\documentclass[12pt]{article} 
\usepackage{epsfig}
\usepackage{amsfonts}
\usepackage{latexsym}
\usepackage{amsmath,amssymb}
\usepackage{verbatim}
\usepackage{mathtools}
\usepackage[dvipdfm,hypertex]{hyperref}
\numberwithin{equation}{section}

 \def\p{\partial}

\newcommand{\bea}{\begin{eqnarray}}
\newcommand{\eea}{\end{eqnarray}}
\newcommand{\be}{\begin{equation}}
\newcommand{\ee}{\end{equation}}
\newcommand{\ba}{\begin{align}}
\newcommand{\ea}{\end{align}}

\newcommand{\W}{\mathcal{W}}

\newcommand{\R}{\mathbb{R}}
\newcommand{\Z}{\mathbb{Z}}

\newcommand\rref[1]{(\ref{#1})}

  \makeatletter
  \let\over=\@@over \let\overwithdelims=\@@overwithdelims
  \let\atop=\@@atop \let\atopwithdelims=\@@atopwithdelims
  \let\above=\@@above \let\abovewithdelims=\@@abovewithdelims

\begin{document}

\begin{titlepage}
\vspace{6cm}
\vfil\

\begin{center}
{\LARGE The Gravitational Exclusion Principle and \\
\vspace{.2cm}
Null States in Anti-de Sitter Space}

\vspace{6mm}

{Alejandra Castro\footnote{e-mail: {\tt acastro@physics.mcgill.ca}}$^{a}$, 
Thomas Hartman\footnote{e-mail: {\tt hartman@ias.edu}}$^{b}$
 \&
Alexander Maloney\footnote{e-mail: {\tt maloney@physics.mcgill.ca}}$^{a}$
}\vspace{6.0mm}\\
\bigskip
{$^a$ \it  Department of Physics, McGill University, Montreal, QC, Canada}\\
{$^b$ \it School of Natural Sciences, Institute for Advanced Study, Princeton, NJ, USA}
\medskip
\vfil

\end{center}
\setcounter{footnote}{0}

\begin{abstract}
\noindent
The holographic principle implies a vast reduction in the number of degrees of freedom of quantum gravity.  This idea can be made precise in AdS${}_3$, where the the stringy or gravitational exclusion principle asserts that certain perturbative excitations are not present in the exact quantum spectrum.  We show that this effect is visible directly in the bulk gravity theory: the norm of the offending linearized state is zero or negative.  When the norm is negative, the theory is signaling its own breakdown as an effective field theory; this provides a perturbative bulk explanation for the stringy exclusion principle. When the norm vanishes the bulk state is null rather than physical. This implies that certain non-trivial diffeomorphisms must be regarded as gauge symmetries rather than spectrum-generating elements of the asymptotic symmetry group. This leads to subtle effects in the computation of one-loop determinants for Einstein gravity, higher spin theories and topologically massive gravity in AdS${}_3$.  In particular, heat kernel methods do not capture the correct spectrum of a theory with null states.
\end{abstract}
\vspace{0.5in}

\end{titlepage}
\newpage

\tableofcontents

\section{Introduction}

The holographic principle asserts that the number of states describing a region of space is proportional to the area of the region in Planck units.  This  constrains the number of degrees of freedom in a theory of quantum gravity and implies that the vast majority of apparently consistent perturbative states are not present in the full quantum theory.   This property is one of the defining features of the AdS/CFT correspondence, in which bulk gravitational degrees of freedom are dual to those of a boundary  theory in one less dimension. One might expect that this reduction in the number of states is  invisible in perturbation theory, as it involves complicated non-perturbative processes in  which excitations collapse to form a black hole. The goal of this paper is to show that this is not always the case.   We will see that for certain theories of gravity many  states are removed from the theory at the perturbative level, and that moreover the number of states removed is in precise accord with the holographic principle.

We will consider three dimensional gravity in Anti-de Sitter space, focusing on Einstein gravity and its higher-spin or supersymmetric generalizations.  Such theories  possess black holes as well as linearized graviton-like states.  These theories behave much like their higher dimension cousins and are expected to exhibit  the same holographic bound on their density of states.  Indeed, various holographic duals to theories of three dimensional gravity are known (see \cite{Kraus:2006wn} for a review).  

The observation that  perturbative degrees of freedom disappear in AdS${}_3$ gravity was first made in the context of supersymmetric string compactifications, where it was dubbed the stringy exclusion principle  \cite{Maldacena:1998bw}.  The idea was revisited  in the context of higher spin theories of gravity without known string theory interpretations, where it was referred to more generally as the gravitational exclusion principle \cite{Castro:2010ce}.   In both cases the observation was based on known features of  conformal field theories and no bulk explanation was given for the disappearance of apparently consistent perturbative gravity states.  

We will show that these exclusion principles have a simple explanation from the point of view of bulk perturbation theory.  The linearized spectrum includes both single and multi-particle states built out of boundary gravitons and their supersymmetric or higher spin generalizations.  The norms of these states can be computed using the Brown-Henneaux procedure \cite{Brown:1986nw}.  In certain cases, we will see that the norm of a multiparticle state is zero or negative.  When the norm is zero the  state is null and should not be considered a physical excitation.  This is the gravity analogue of a well known feature of CFTs, namely the appearance of null vectors  in representations of the conformal group.   The novelty is that  this logic can be applied directly to the bulk gravity theory.

In  cases relevant to the stringy exclusion principle the norm of a perturbative state becomes negative.  This does not signal a sickness of the theory but rather a breakdown of low-energy effective field theory.  This feature is common in non-gravitational quantum field theories; a nonrenormalizable effective theory will produce unphysical, unitarity violating scattering amplitudes at energies above the cutoff.  The multiparticle states which violate the stringy exclusion principle appear only if we use an effective field theory (low energy supergravity) outside of its regime of validity.  

The question of how non-perturbative effects are visible in perturbation theory -- and how the breakdown of low energy effective field theory manifests itself -- is one of the fundamental questions in quantum gravity.   A full answer to this question would explain the apparent non-unitary of Hawking evaporation, which is based on an application of effective field theory in the presence of horizons.   
We can now see in a straightforward manner how the low energy effective theory breaks down in certain circumstances in AdS$_3$, providing a partial answer to this question: it breaks down through the appearance of zero or negative norm multiparticle states.

We will describe the mechanism underlying null and negative norm states in detail in section 2.  The simplest case where this effect arises is  pure Einstein gravity in a quantum mechanical regime where the AdS radius is of order one in Planck units. 
In this regime the theory is truly quantum mechanical and perturbation theory breaks down.  Nevertheless, the theory still has boundary graviton states which are constructed in the usual manner. Our results can be viewed as statements about the classical phase space and symplectic structure in this regime, which would be the starting point for the quantization of the theory of pure gravity.\footnote{In this paper we  focus on the boundary graviton sector as it provides a simple illustration of the mechanism by which states are removed from the spectrum.  Non-perturbative statements are possible as well; we leave this for  future  work \cite{castroetal}.}  In this case the construction mirrors that of minimal model CFTs.  We will show that the standard one-loop determinant is replaced by a minimal model vacuum character, and that the consequent reduction in the number of states is precisely what is needed in order for the theory to obey a suitable holographic bound.  Moreover, we will show that the theory is unitary only if the central charge takes one of the usual allowed minimal model values.

Perhaps a more important application of this mechanism is to the higher spin theories of gravity discussed in \cite{Blencowe:1988gj,Henneaux:2010xg, Campoleoni:2010zq, Gaberdiel:2010pz, Gaberdiel:2010ar}.  This family of theories possesses massless gauge fields of spin $2,\dots,N$ for any $N$.  In this case the exclusion principle leads to null states when the central charge is $c<N-1$, so that for sufficiently large $N$ the central charge can be large. Although the theory appears semi-classical, the perturbative exclusion mechanism described above continues to apply. We will see that many states are projected out and that the central charge must be quantized in order for the spectrum to be unitary.  This resolves a puzzle in the literature,  namely that the one-loop determinant computed using standard heat-kernel methods  \cite{Gaberdiel:2010ar} conflicts with established results involving black hole physics \cite{Castro:2010ce}.  The reason is that the determinant of \cite{Gaberdiel:2010ar} includes null states as well as physical states.  The contradiction disappears once  null states are removed. 

Indeed, recently a number of  computations of one-loop determinants in AdS gravity have led to an improved understanding of the AdS-CFT correspondence \cite{Giombi:2008vd, David:2009xg, Gaberdiel:2010ar}.  The heat kernel methods used in these computations give mathematically correct statements about functional determinants.  However, in some cases they are simply not computing the physically relevant quantity.  These one-loop determinants correctly encode the spectrum for states with a small number of particles, the usual case of interest.  But they fail  to correctly describe  states with a large number of particles where the stringy or gravitational exclusion principle applies.  Simply put, the heat kernel correctly computes the one-loop determinant of the wrong theory, a nonunitary theory which has zero norm states.  These null states should be removed from the correct one-loop determinant.  This can be compared to a similar situation in nonabelian gauge theory: if one-loop diagrams in Yang-Mills are computed without properly accounting for unphysical states by including the Fadeev-Popov ghosts, then the resulting scattering amplitudes will be non-unitary.

This paper is organized as follows.  In section 2 we describe the  computation of the norm of perturbative states in AdS$_3$ using the Brown-Henneaux procedure.  In section 3 these results are applied to Einstein gravity, and it is found that gravity with $c<1$ has boundary gravitons with zero or negative norm.  In section 4, several other applications are described.  First, the results for Einstein gravity with $c<1$ are extended to higher spin gravity with $c<N-1$, where $N$ is the maximal spin. Second, the computation of norms is applied to $N=2$ supergravity, and it is shown that the stringy exclusion principle can be understood as the appearance of negative norm states. Third, null states are discussed in the context of topologically massive gravity at the chiral point where $c_L = 0$.

\section{Norms in AdS${}_3$ Gravity}

We now describe the perturbative approach to the gravitational exclusion mechanism.  We begin with a general discussion of norms in Einstein gravity in three dimensions before moving on to examples. 

Gravity in three dimensional AdS space is, according to the AdS/CFT correspondence, expected to be dual to a two dimensional CFT with  central charge
\be\label{cis}
c = {3 \ell \over 2 G}~.
\ee
Here $\ell$ is the AdS radius and $G$ is Newton's constant.  This applies to Einstein gravity with arbitrary matter content, as long as appropriate boundary conditions are imposed on matter fields. We will first recall the construction of perturbative bulk states in AdS${}_3$, before proceeding to compute their norm.  

The definition of AdS gravity requires a choice of boundary conditions.  With  standard Brown-Henneaux boundary conditions \cite{Brown:1986nw}, only those diffeomorphisms which vanish sufficiently quickly at infinity are true gauge symmetries, in the sense that metrics related by such a diffeomorphism  describe the same state.  Diffeomorphisms which do not vanish at the boundary of AdS are not gauge symmetries but instead act non-trivially on the spectrum of the theory.  

We will denote by $\zeta$ a vector which generates such a symmetry.   
One can then consider the conserved charge $H[\zeta]$ associated with $\zeta$.  This charge is the on-shell value of the gravitational Hamiltonian which generates via Dirac brackets the action of the diffeomorphism $\zeta$.  The achievement of Brown \& Henneaux was to show that the generators with finite charge generate a copy of the two dimensional conformal group with central charge $c$ given by \rref{cis}.

We now examine the spectrum of gravitons around empty AdS${}_3$.  
Perturbative states are obtained by acting on the empty AdS${}_3$ ground state with infinitesimal diffeomorphisms $\zeta$.  
Being diffeomorphisms applied to the original metric, such states are clearly solutions to the equations of motion.  The novelty is that the state is now a genuine physical state, rather than  pure gauge as one might expect, 
because it carries a conserved charge.  
As their presence in the physical spectrum is a consequence of the boundary conditions, such states are known as a boundary gravitons.  

We now compute the norm of a boundary graviton state.    
We start with the standard expression for the Klein-Gordon inner product between two linearized metric fluctuations
\be
(\delta_1 g , \delta_2 g) \equiv \int_{\Sigma} \omega(g, \delta_1 g^* , \delta_2 g) \ .
\ee
Here the integration is over a spatial slice, $\delta_{1,2}g$ are linearized solutions to the equations of motion expanded around a background metric $g$, and $\omega$ is the symplectic current. This is the standard expression for the norm of a linearized excitation; for example the symplectic current of a scalar field reproduces the usual Klein-Gordon inner product, $||\phi||^2 = \int \phi^* \overset{\leftrightarrow}{\partial}_t \phi$. In general relativity, the symplectic current  $\omega$ can be computed following \cite{Crnkovic:1986ex}.  In the present case, however, the explicit expression for $\omega$ is  subtle as it includes contributions from boundary terms in the action.

Thankfully, an explicit computation of $\omega$ is not necessary.  We simply note that the variation of the charge $H[\zeta]$ under the metric perturbation $\delta g$ is also defined in terms of the symplectic current by
\be\label{defham}
\delta H[\zeta] = \int_\Sigma \omega(g, \delta g, \delta_\zeta g) \ .
\ee
From this it follows that the norm of the boundary graviton state created by the diffeomorphism $\zeta$ is
\be
||\zeta||^2 = \delta_{\zeta^*} H[\zeta] \ ,
\ee
The defining property of the charge $H[\zeta]$ is that it is a generator of the diffeomorphism $\zeta$.  Thus the norm is
\be\label{normcom} 
||\zeta||^2 = \{ \ H[\zeta^*] \ , \ H[\zeta] \ \}_{D.B.} \ .
\ee
where $\{, \}_{D. B.}$ is the Dirac bracket.  In the quantization of the theory this Dirac bracket will be promoted to a commutator.   We note that this formula requires a choice of background metric $g$ which in this case is empty AdS${}_3$.

We conclude that the norm of the boundary graviton states can be computed from the Dirac bracket algebra of charges.  This expression will properly account for boundary contributions to the action.  The result (\ref{normcom}), though simple, is not widely recognized. In the rest of this paper we will explore its applications to several examples of AdS$_3$ gravity.

\section{Null States in Einstein Gravity}

The simplest  application of (\ref{normcom}) is to Einstein gravity with $c <1$.  In this regime the  theory is truly quantum mechanical and there is no clear distinction between perturbative and non-perturbative effects.   In this paper we focus only on the boundary graviton sector of the theory; we will show that there are multiparticle states with zero or negative Klein-Gordon norm. 
The quantum theory of pure gravity, if it exists, should involve the quantization of a classical phase space with the symplectic structure implied by the Einstein Hamiltonian.  The quantization of the boundary graviton sector  is a necessary first step towards the quantization of three dimensional gravity in this regime.

Let us now describe the boundary gravitons a bit more explicitly.  In  coordinates
\be
ds^2 = d\rho^2 -\cosh^2 \rho d\tau^2 + \sinh^2 \rho d\phi^2 ~,
\ee
the vectors $\zeta$ can be expanded in a basis
\be\label{zetadef}
\zeta_n = e^{i n u}\left(\p_u + n^2 e^{-2 \rho} \p_v - i {n\over 2 } \p_\rho \right) + \dots ,~~~~~{\bar \zeta_n} = e^{i n v}\left(\p_v + n^2 e^{-2 \rho} \p_u - i {n\over 2 } \p_\rho \right) + \dots ~,
\ee
where $u=\tau+\phi$, $v=\tau-\phi$ and `$\dots$' denotes terms which do not contribute to the charges $H[\zeta]$. We denote the corresponding charges by $L_n = H[\zeta_n], {\bar L}_n = H[{\bar \zeta}_n]$.  They obey the Virasoro algebra
\be\label{virasoro}
[L_n,L_m] = (n-m)L_{n+m}+{c\over 12} n(n^2-1)\delta_{n+m, 0}~,
\ee  
with similar expressions for ${\bar L}_n$.  

We denote by $|0\rangle$ the empty AdS vacuum state.  This state has zero energy and angular momentum, hence will be annihilated by the operators $L_0$ and ${\bar L}_0$.  Moreover, as it is the state with lowest energy it will be annihilated by all of the lowering Virasoro operators $L_{n}, {\bar L}_{n}$ with $n >0$ which decrease $L_0$ and $\bar L_0$.    Thus it is a weight-zero primary\footnote{We are using the ``plane" normalization, where the vacuum is taken to have weight zero.  This is the natural normalization if we think of the vacuum as obtained by radial quantization on the plane with the identity operator (which has trivial scaling dimension) inserted at the origin.  This differs from the ``cylinder" normalization by a shift by the conformal anomaly term $c/24$.} 
\be\label{primary}
L_n |0\rangle =0~~~~n\ge0~.
\ee
Boundary gravitons  are obtained by applying the Virasoro raising operators $L_{-n}, n >0$.  For example, the single-particle states  are obtained by applying a diffeomorphism to the vacuum
 \be
 L_{-n}|0\rangle~~~~~n>0~.
 \ee 
Multi-particle states are obtained by applying multiple infinitesimal diffeomorphisms 
\be\label{verma}
L_{-n_1}\dots L_{-n_k}|0\rangle,~~~~~n_i>0~.
\ee
In the CFT language these states are Virasoro descendants of the vacuum, and the space of all such states is the vacuum Verma module, but we need not make any reference to holography to describe the perturbative bulk spectrum in this way.  For large values of the central charge the states \rref{verma} are legitimate perturbative states.  However, when $c<1$ the structure of Virasoro representations is more constrained, and it is for this reason that some of the states \rref{verma} must disappear from the spectrum.

To see this we must compute the norm of  these states using \rref{normcom}.  We first note that, from \rref{zetadef}, $H[\zeta_n^*] = L_{-n}$.  This is the usual statement that
\be\label{adjoint}
L_{n}^\dagger = L_{-n} \ .
\ee
Here $\dagger$ denotes the adjoint with respect to the bulk inner product defined by \rref{normcom}.
Given this, along with the algebra \rref{virasoro}, we can compute the norm of the boundary graviton states.  The computation of the norm of is identical to the CFT computation, and in particular the existence of null vectors when $c<1$ mirrors precisely the discussion which arises in the construction of minimal models.  We will summarize this discussion for completeness, focusing on the gravitational interpretation.

Let us start with the state $L_{-1} |0\rangle$, which has zero norm
\be\label{simplenorm}
|| L_{-1} |0\rangle ||^2 = \langle 0 | L_{1} L_{-1} |0\rangle = \langle 0 | [L_1, L_{-1}] |0\rangle=0 \ .
\ee
This is a consequence of the fact that, because the empty AdS ground state has $so(2,2) = sl(2,\R)\times sl(2,\R)$ symmetry, it is annihilated by the generator $L_{-1}$ in $sl(2,\R)$.  Thus even though $L_{-1}$ is an element of the asymptotic symmetry group it does not lead to a physical state when acting on the ground state.  The state has zero norm and  must be dropped from the spectrum in order to obtain a unitary theory. Note that although we are using notation familiar from conformal field theory, (\ref{simplenorm}) refers only to bulk quantities.  On the left-hand side is a Klein-Gordon norm, and the notation $\langle 0 | [L_1, L_{-1}] |0\rangle$ denotes the Dirac bracket of two boundary gravitons, evaluated in the pure AdS$_3$ background.

One can now ask whether the same thing happens with other, more complicated boundary graviton states.  The boundary graviton norms are straightforward to compute.  In general, the answer is a function of the central charge:
\begin{itemize}
\item When $c>1$  all boundary graviton states have positive norm.
\item When $c<1$ some boundary graviton states have non-positive norm.  There will be a state with negative norm unless 
\be
c = 1-{6\over p(p+1)}~,
\ee
for some integer $p> 2$.  In this case some of the boundary graviton states have zero norm.
\end{itemize}
The proof of these facts follow in a straightforward way from the Kac determinant formula; we refer the reader to \cite{DiFrancesco:1997nk} for a review.  
 
Rather than repeating the details of this proof, let us illustrate the idea with the following simple example.  Consider the case $p=3$, i.e. $c={1\over 2}$.  In this case the first nontrivial null descendant is 
$ \chi_{-6}|0\rangle $ where
\be\label{chi6}
\chi_{-6} \equiv L_{-6} + {22\over 9} L_{-4}L_{-2} - {31\over 36} L_{-3}^2 - {16\over 27}L_{-2}^3 \ .
\ee
This can be seen by the straightforward computation
\be\label{chinorm}
 ||  \chi_{-6} |0 \rangle ||^2 = \langle 0\,|\,[\chi_6 \ , \ \chi_{-6}]\,|\,0\rangle = 0 \ .
\ee
It is important to note that the state $\chi_{-6}|0\rangle$ is non-linear in the charges, thus it should not be regarded as being created by a single diffeomorphism; it is a multiparticle state. The commutator in (\ref{chinorm}) is a true commutator in the perturbative quantum theory, not a Dirac bracket, but it is defined in the usual way by promoting the single particle Dirac brackets to quantum commutators.

Of course, $\chi_{-6}|0\rangle$ is but the first of many such null states; we will count the number of such states below and demonstrate that the vast majority of boundary graviton states  are null.  Just like $L_{-1} |0\rangle$ these states will not appear in the physical spectrum of the theory.  One can regard these new null states  as generators of an additional gauge symmetry which is  present only in the quantum $c<1$ regime.

Finally, we emphasize that in this entire discussion we have used here only the Virasoro algebra \rref{virasoro}, the fact that the vacuum is a primary of weight zero \rref{primary} and the expression for the adjoint \rref{adjoint}.  All of these properties were understood directly from the gravity point of view, so although we have used the CFT language -- and used results from the CFT literature -- this discussion should be regarded as a direct gravity proof that perturbative states are removed from the spectrum.

\subsection{The One-loop Determinant}

We would now like to count more precisely the number of boundary graviton states.  We will do so by computing the partition function 
\be
Z^{vac}(\tau)  = {\rm Tr} \, q^{L_0},~~~~~~q=e^{2\pi i \tau}~,
\ee
where the trace is over all physical (positive norm) boundary graviton states.
The parameter $\tau$, which appears in this expression as a formal expansion parameter, can be regarded both as the (complexified) temperature as well as the conformal structure parameter of a torus at the boundary of Euclidean Anti-de Sitter space.
We have written here only the holomorphic part of the partition function; of course one can define an identical anti-holomorphic partition function ${\bar Z}^{vac} (\bar \tau)= {\rm Tr} \,{\bar q}^{{\bar L}_0}$.  The full  one-loop contribution to the AdS${}_3$ partition function is $|Z^{vac}|^2$.   

Let us first consider the case where $c>1$.  Then $Z^{ vac}$ is the trace over the full Verma module with only the constraint that $L_{-1}|0\rangle$ is null.  The result is\footnote{For future convenience, we adopt here  the ``cylinder" normalization where the ground state  has dimension $-c/24$ rather than zero.  This leads to the prefactor $q^{-c/24}$.  } 
\be\label{Zpert}
Z^{vac}_{c>1} = q^{-c/24}\prod_{n=2}^{\infty} {1\over 1-q^n}  = q^{-c/24}\left(1 + q^2 + q^3 + 2q^4 +2q^5 + 4q^6 + \cdots\right)~.
\ee
This computation of the one-loop determinant by counting boundary gravity states was discussed in \cite{Witten:2007kt, Maloney:2007ud}.
The result agrees with direct computations of the one-loop partition function around AdS evaluated using  heat kernel methods \cite{Giombi:2008vd, David:2009xg}.

For theories with $c<1$ the answer involves a trace over a more complicated representation of the Virasoro algebra.  
For example, when $c=1/2$ the result is 
\be
Z^{vac}_{c=1/2} = q^{-c/24}\left(1 + q^2 + q^3 + 2q^4 + 2q^5 + 3q^6 + \cdots\right) ~.
\ee
The difference between these last two formulas is precisely the absence of $\chi_{-6}|0\rangle$ in the spectrum.

In general, $Z$ is equal to the vacuum character of a minimal model CFT, usually denoted $\chi_{1,1}$
\be
Z^{vac}_{c<1}(\tau) = \chi_{1,1}(\tau)~.
\ee 
Explicit expressions for this character are reviewed in \cite{DiFrancesco:1997nk}.  It is 
\begin{align}\label{chi11}
\chi_{1,1} = q^{-c/24}\left(\prod_{n=2}^{\infty} {1\over 1-q^n}\right)
\left(1+\sum_{k=1}^\infty (-1)^k \left( q^{h_{1+k(p+1),(-1)^k +(1-(-1)^k)p/2}}\right.\right.+\\ \left.\left.q^{h_{1,kp+(-1)^k +(1-(-1)^k)p/2}}\right)\right)
\end{align}
where \be\label{hrs}
h_{r,s} = {((p+1)r-ps)^2-1\over 4 p(p+1)}~.
\ee

It is important to emphasize that this result differs from the usual expression for the one-loop partition function as a ratio of functional determinants.   The functional operators appearing in these determinants are obtained by linearizing the equations of motion for the metric and ghost fields.   The resulting determinant properly computes the trace over the Hilbert space of small fluctuations only if one assumes that multi-particle states are built out of single-particle states in the usual manner.  This assumption is of course valid in the semi-classical limit.  In certain cases, however, this assumption is false.  For example, we have seen that for Einstein gravity in a quantum regime the norm of multi-boundary graviton states can be zero or negative even though the single particle states all have positive norm.  This is simply not accounted for in the standard one-loop computation typically done using heat kernel methods.  In principle a BRST procedure could be used to properly account for the new null states.  In the present case, however, this is not necessary.  The one-loop determinant \rref{chi11} can be found  by directly enumerating boundary gravitons states with positive norm.

\subsection{Implications for Holography}

 We conclude the discussion of pure gravity by noting that the mechanism described above removes precisely the number of states  needed in order to satisfy the holographic bound.  Three dimensional AdS gravity possesses black hole solutions \cite{Banados:1992wn}.  The entropy of a black hole is proportional to its horizon area, and in fact matches the asymptotic density of states of a conformal field theory given by Cardy's formula \cite{Strominger:1997eq}
 \be\label{sbh}
 S_{BH} = {A\over 4G} = {2 \pi }\sqrt{{c\over 6} L_0}+ 2\pi \sqrt{{c\over 6}\bar L_0}~.
 \ee
Here we have expressed the area as a function of the mass  $L_0+{\bar L}_0$ and  angular momentum $L_0 - \bar L_0$.  This entropy counts the number of black hole microstates, which under normal circumstances vastly exceeds the number of graviton states in the theory.
 
However, if $c$ is of order one this is no longer the case.  Indeed, if one estimates the number of boundary graviton states counted by \rref{Zpert} one finds the following asymptotic growth at high order\footnote{This is computed by using the modular transformation properties of the eta function $\eta(\tau) = q^{1/24}\prod_{n=1}^\infty (1-q^n)$.  The derivation is essentially identical to that of the Cardy or Hardy-Ramanujan formula.}
\be\label{spert}
S_{graviton} =2 \pi \sqrt{ {L_0\over 6}}+ 2\pi \sqrt{{\bar L_0\over 6}}~.
\ee
 Comparing with \rref{sbh} we see that this growth in the density of states  conflicts with the holographic expectation that  black hole entropy places a fundamental upper bound on the number of degrees of freedom.
 
On the other hand, we have seen that in fact many of the states which contribute to the entropy \rref{spert} are not present in the spectrum.  Instead we must compute the density of states from the vacuum character $\chi_{1,1}$ in \rref{chi11}.  In particular, we have 
\be
S = \log N(L_0) + \log N({\bar L_0})~,
\ee
where the number of states is given by the inverse Laplace transform of $\chi_{1,1}$
\be\label{nisbb}
N(L_0) = {1\over 2 \pi i} \int_{\cal C} {\chi_{1,1}(q) \over q^{L_0+1}} dq~.
\ee 
Here $\cal C$ is a contour which encloses the origin.     To compute this we will use known modular transformation properties of minimal model characters. 

In order to estimate $N(L_0)$ we will choose a contour which approaches $|q|=1$ where $\chi_{1,1}$ can be approximated by elementary functions and the integral evaluated.  This is the standard approach used to derive the Cardy formula.    In the present case, however (unlike the usual application of the Cardy formula) $\chi_{1,1}(\tau)$ is not modular invariant.  Rather it transforms in a vector representation of the modular group.  For example, under the $S$ transform $\tau \to -1/\tau$ we have 
\be\label{chis}
\chi_{1,1}(\tau) = -2 \sqrt{2 \over p(p+1)}\sin\left({p \over p+1}\pi \right)\sin\left({p+1 \over p}\pi \right)\chi_{1,1}(-1/\tau) + \dots~,
\ee
where $\dots$ denotes other minimal model characters.  We are only interested in the behaviour where $|q|\to 1$ and $\tau \to 0$.  The other minimal model characters are subleading in this limit, as they correspond to states with higher (i.e. non-zero) weight. From \rref{chi11} we see that
\be\label{chisaddle}
\chi_{1,1} (-1/\tau) \sim e^{-2 \pi i {c \over 24}{1\over \tau}}~,~~~~~\tau \to 0~,
\ee
so may compute the integral \rref{nisbb} to obtain
\be\label{nfinal}
N(L_0) \sim e^{2 \pi\sqrt{ { c L_0\over 6}}+ 2\pi\sqrt{{c \bar L_0\over 6}} }~.
\ee
Here we have used the saddle point approximation and neglected all constants and power law terms.

We note that this  precisely saturates the holographic bound \rref{sbh}.  This means that we have lost exactly the number of perturbative states required by holography.
From the boundary point of view this statement can be understood as a consequence of the fact that any unitary CFT will obey Cardy's formula.

\section{Applications to Various Theories of 3d Gravity}

We now turn to other applications of our expression for the norm of a boundary graviton state (\ref{normcom}).  We begin by describing null states in higher spin gravity.  This generalizes the previous discussion to theories which possess  extended chiral symmetry algebras.  We then describe the appearance of negative norm states in supergravity theories in AdS${}_3$ coming from string compactifications; these negative norm states signal the breakdown of effective field theory due to the stringy exclusion principle.  Finally, we discuss chiral gravity.

\subsection{Higher Spin Gravity}

The argument described in section 3 applies to certain higher spin theories of AdS${}_3$ gravity, such as those described in \cite{Aragone:1983sz,Blencowe:1988gj,Bergshoeff:1989ns}.  These theories are most clearly described using the Chern-Simons formulation of 3D gravity, where the usual $SO(2,2)=SL(2,\R)\times SL(2,\R)$  gauge group is replaced by $SL(N,\R) \times SL(N,\R)$.  In this case the algebra of asymptotic symmetries is enhanced from two copies of the Virasoro algebra to two copies of ${\cal W}_N$.  
For more information on these higher spin algebras and their applications see \cite{Bouwknegt:1992wg}.

These algebras are centrally extended, as in the pure gravity case, with central charge \cite{Henneaux:2010xg,Campoleoni:2010zq}
\be
c= {3 \ell \over 2G} ~.
\ee
With the replacement of the Virasoro algebra by ${\cal W}_N$ the details of the previous argument carry over almost immediately.  In this case the perturbative states include not just boundary gravitons but higher spin versions of the boundary gravitons where a non-trivial ${\cal W}_N$ generator is applied to the vacuum.

As before, the norm of a boundary graviton state follows from a simple algebraic computation.  The computation mirrors precisely that which arises in the construction of ${\cal W}_N$ minimal models, as reviewed in e.g. \cite{Bouwknegt:1992wg, Beltaos:2010ka}.  The primary subtlety is that the $\W_N$ algebra is not a standard Lie algebra. The algebra is nonlinear in the sense that the commutator of two generators is a polynomial in the other generators.   This creates an important distinction between the classical and quantum algebra.  As shown in \cite{Henneaux:2010xg,Campoleoni:2010zq,Gaberdiel:2011wb,Campoleoni:2011hg} the algebra of higher spin gauge transformations in the bulk leads to a classical ${\cal W}_N$ algebra.  Upon quantization operator products must be regulated using an appropriate normal ordering prescription; this gives the corresponding quantum $\W_N$ algebra.

It is useful to illustrate this in the simple example of $N=3$, where commutation relations are relatively simple and the representation theory can be worked out explicitly.  The quantum algebra is
\bea
[L_{n},L_{m}]&=&(n-m)L_{n+m}+{c\over 12}(n^3-n)\delta_{m+n,0}\cr
[L_{n},W_{m}]&=&({2n}-m)W_{n+m}\cr
[W_n,W_m]&=&{c\over 360}n(n^2-1)(n^2-4)\delta_{m+n,0}+{16\over 22+5c}(n-m)\Lambda_{m+n}\cr
&&+(n-m)\left({1\over 15}(m+n+3)(m+n+2)-{1\over 6}(m+2)(n+2)\right)L_{m+n}
\eea
with
\be
\Lambda_{m}=\sum_{q\in\Z}:L_{m-q}L_{q}:-{3\over 10}(m+3)(m+2)L_{m}~.
\ee
Here we use the notation $: \,:$ to denote annihilation-creation normal ordering. 

The vacuum character coming from the Verma module of $\W_3$  is obtained by enumerating $\W_3$ descendant states.   
As in the case of Einstein gravity, the vacuum is annihilated by the generators of the rigid $sl(3)$ subgroup of  the chiral algebra:
\be
W_{-2}|0\rangle=W_{-1}|0\rangle = L_{-1}|0\rangle=0~.
\ee
This reflects the fact that the $AdS_3$ ground state is an $sl(3)\times sl(3)$ invariant primary state.
Accounting for this, the vacuum Verma module character is  (in plane normalization)
\bea\label{X3}
\chi^{(3)}&=&\prod_{s=2}^3\prod_{n=s}^\infty(1-q^n)^{-1}\cr
&=&1+q^2+2q^3+3q^4+4q^5+8q^6+\cdots
\eea
This result matches the  heat kernel computation of the one-loop determinant for higher spin gravity \cite{Gaberdiel:2010ar}.

This  Verma module may include states of non-positive norm.  The structure of the $\W_3$ representations was described in \cite{Fateev:1987vh}.
The result is that
\begin{itemize}
\item When $c>2$  all of the states have positive norm
\item When $c<2$ some states have non-positive norm.  There will be a state with negative norm unless 
\be\label{wminthree}
c = 2\left(1-{12\over p(p+1)}\right)~,
\ee
for some integer $p> 3$.
\end{itemize}
Just as in the pure gravity case, the result \rref{X3} must be modified for small values of the central charge.

It is possible to write explicit expressions for the vacuum character, analogous to \rref{chi11}, but we will not do so  here. Instead we will illustrate this idea for the simplest nontrivial case, $c=4/5$  ($p=4$ in \rref{wminthree}), which is the 3-state Potts model.   The ${\cal W}_3$ vacuum character is not given by \eqref{X3} but rather by \cite{DiFrancesco:1997nk}
\bea\label{chipotts}
\chi^{(3)}_{c=4/5}&=&\chi_{1,1}+\chi_{4,1}\cr
&=&1+q^2+2q^3+3q^4+4q^5+7q^6+\cdots
\eea
Here $\chi_{1,1}$ and $\chi_{4,1}$ are standard (Virasoro) minimal model characters for the 3-state Potts model.
Comparing \rref{chipotts} with \rref{X3} we see that the first null state appears at level six.   The null state is given by a linear combination of the generators \be
L_{-6}, L_{-4}L_{-2},(L_{-3})^2,(L_{-2})^3,W_{-6}, W_{-4}L_{-2}, (W_{-3})^2~,\ee as can be verified by explicit computation.  We emphasize that this null state is special to the $\W_3$ algebra; by itself, the Virasoro algebra with $c=4/5$ does not have a null state at level six.  Of course this is but the first of an infinite number of null states. 

We now consider the generalization of this argument to $\W_N$ algebras for larger values of $N$.  In this case the representation theory is more complicated and it is more difficult to make exact statements.  However, there is considerable evidence for the following two statements\footnote{The second statement was proven (in the CFT language) in \cite{Fateev:1987zh}.  The first statement is consequence of the conjecture on p. 100 of  \cite{Bouwknegt:1992wg}, where strong evidence was presented  in its favour.}
\begin{itemize}
\item When $c>N-1$ all of the boundary graviton states and their higher spin generalizations  have positive norm.
\item When $c<N-1$ some states have non-positive norm.  When
\be\label{wmin}
c = (N-1)\left(1-{N(N+1)\over p(p+1)}\right)
\ee
 for some $p > N$,
these states are null and no states have negative norm.
\end{itemize}
We emphasize that when $N$ is large, null vectors can appear even when the central charge is large, as long as $c<N-1$.   Thus the gravitational exclusion principle applies even to AdS spaces whose size is large in Planck units.  In this case, even though the theory
 appears to have 
 a semi-classical description,  the standard expression for the one-loop partition function in terms of functional determinants fails to account for the appearance of null-states. The correct determinant can be computed only by enumerating states with strictly positive norm.    

We can also show that this mechanism removes precisely the number of states required by the holographic bound.  This is a simple generalization of the argument of section 3.2.     The number of states in the vacuum Verma module of $\W_N$  grows like
\be\label{}
S_{Verma} = 2 \pi \sqrt{(N-1) L_0 \over 6}+  2 \pi \sqrt{(N-1) {\bar  L_0} \over 6}
\ee for large $L_0$ and $\bar L_0$.  When $c<N-1$ this exceeds the black hole entropy \rref{sbh}, as was noted in \cite{Castro:2010ce}.  We now see the resolution of this puzzle.  The correct number of positive norm states is given by the contour integral 
\be\label{nis}
N(L_0) = {1\over 2 \pi i} \int_{\cal C} {\chi^{(N)}(q) \over q^{L_0+1}} dq~
\ee 
where $\chi^{(N)}$ is the $\W_N$ minimal model vacuum character.  Explicit expressions for such characters are rather complicated, but fortunately such expressions are not necessary.  We can simply use the fact that this character transforms in a finite dimensional representation of the modular group, so that 
\be\label{chins}
\chi^{(N)}(\tau) =  n \chi^{(N)}(-1/\tau) + \dots
\ee
where $n$ is independent of $\tau$.  Here $\dots$ denotes contributions  from other minimal model characters which vanish in the $\tau \to 0$ limit.    Using a saddle point approximation as in \rref{chisaddle} leads to the result \rref{nfinal} for the number of states.  This precisely saturates the holographic bound.

\subsection{Stringy Exclusion Principle}
The most well-known case where quantum gravity effects remove perturbative states is the stringy exclusion principle \cite{Maldacena:1998bw}. 
Although the stringy exclusion principle first arose in considerations of type II string theory on orbifolds of AdS$_3\times S^3\times M^4$ it can be regarded as a general feature $N=2$ supergravity in AdS${}_3$.   
 For the sake of simplicity, we will therefore restrict our discussion to AdS$_3$ supergravity.  
The essential observation is that in a CFT with $N=2$ superconformal symmetry the  $R$-charges of chiral primaries  are bounded by the central charge \cite{Lerche:1989uy}
\be\label{excl}
q \leq {c\over 3} \  .
\ee
In the AdS bulk, chiral primaries are described by 
weakly interacting
fields.   Thus it is possible to form multiparticle states which are also chiral primaries.  A multiparticle state with enough particles will eventually violate (\ref{excl}), and hence cannot be present in the exact quantum spectrum of the theory.  This effect is typically assumed to be invisible in perturbative 
supergravity.
 
We  now argue that the necessity of the stringy exclusion principle is visible in bulk perturbation theory.\footnote{
This argument is similar to (but simpler than) that of \cite{Evans:1998qu,Evans:1998wq}, who showed that the stringy exclusion principle is related to the no-ghost theorem for certain string worldsheet theories on AdS${}_3$.}  
We start by identifying the physical states of the bulk theory.  The supersymmetric version of the Brown-Henneaux argument gives the super-Virasoro algebra \cite{Banados:1998pi,Henneaux:1999ib}
\bea\label{n2al}
[L_{n},L_{m}]&=&(n-m)L_{n+m}+{c\over 12}(n^3-n)\delta_{m+n,0}\cr
[L_{n},G_{r}^\pm]&=&({n\over 2}-r)G_{n+r}^\pm\cr
\{G_{r}^-,G_{s}^+\}&=&2L_{r+s}-(r-s)J_{r+s}+{c\over 3} r^2\delta_{r+s,0}
\eea
Here $n\in \Z$ and $r$ is a half-integer. The $U(1)$ current algebra is
\bea
[L_m,J_n]&=&-nJ_{m+n}\cr
[J_m,J_n]&=& {k\over 2}m \delta_{m+n,0}\cr
 [J_n,G^{\pm}_r]&=&\pm G_{n+r}^{\pm}
\eea
Supersymmetry implies that the level is given by $k=c/6$.  As before, we have promoted  Dirac brackets to quantum commutators.  

Consider a bulk state $|\phi\rangle$ which is a chiral primary of the algebra \rref{n2al} of dimension $h=q/2$ and charge $q$.   A multi-particle state $|\phi_{(N)}\rangle$ consisting of $N$ identical such excitations can be constructed by taking the $N$th power of $|\phi\rangle$. It follows from the algebra that 
\be
|| G_{-3/2}^+ |\phi_{(N)}\rangle ||^2 = \langle \phi_{(N)} |2L_0 - 3 J_0 + 2c/3 |\phi_{(N)}\rangle  = 2(c/3 - qN)~.   
\ee
 If $N$ exceeds $c$ then this descendant state has negative norm when $qN>c/3$.  This signals a breakdown in 
the bulk low-energy effective theory
precisely at the point required by the stringy exclusion principle.  

We emphasize that even though the need for a stringy exclusion principle is visible in perturbation theory -- otherwise the theory would be non-unitary -- the actual mechanism by which these states are removed from the spectrum is non-perturbative.  Thus the situation is rather different from that of pure gravity or a higher spin gauge theory where states become null and simply drop out of the physical spectrum for certain values of the central charge.  In the present case the perturbative states have negative norm and  are removed only by non-perturbative effects.  In particular,  for sufficiently large $N$ a multiparticle state $|\phi_{(N)}\rangle$ will backreact on the geometry.  Including this backreaction the state will then be  described by one of the geometries of Lunin, Maldacena and Maoz \cite{Lunin:2002iz}.  
We expect that such geometries with $qN>c/3$, if they exist, will have an instability such as a naked singularity or closed timelike curve.  It would be nice to verify this.

\subsection{Chiral Gravity}

We finally comment on the case of three dimensional gravity with a gravitational Chern-Simons term.  This theory is usually referred to as topologically massive gravity\cite{Deser:1981wh}.  In this case the theory is third order in derivatives and parity odd so that the left and right-moving central charges are not equal. At the chiral point, $c_L=0$, the computation of linearized fluctuations of the metric is rather subtle: the spectrum of the theory and its charges $H[\zeta]$  depend on the fall-off conditions of the metric at the boundary. Two interesting scenarios are\footnote{See e.g. \cite{Maloney:2009ck} and citations therein for a more detailed discussion of these boundary conditions.}
\begin{itemize}
\item Chiral gravity, in which Brown-Henneaux boundary conditions are imposed \cite{Li:2008dq}. The linearized spectrum consists of only right-moving excitations.  Left-moving diffeomorphisms are pure gauge.
\item Log gravity, where  the boundary conditions allow for a logarithmically growing  mode that has finite norm \cite{Grumiller:2008qz}. 
\end{itemize}
It is important to emphasize that these different choices of boundary conditions represent different definitions of the theory
with different Hamiltonians.
These two different theories will have different physical content both at the linear and non-linear level.

The one-loop determinants for chiral gravity and log gravity will be quite different.  In chiral gravity, the left-moving charges $H[\zeta_L]$ vanish, and  thus the norms of the corresponding boundary graviton states vanish as well according to \rref{normcom}.  This was the essential idea of the proof of the chiral gravity conjecture of \cite{Strominger:2008dp}, which provides an additional simple illustration of the phenomenon of gravitational exclusion.   The one-loop determinant is then precisely the holomorphic part of the pure Einstein gravity determinant, as computed in \cite{Maloney:2009ck}.

In log gravity the situation is different. The boundary conditions lead to zero norm states that violate unitarity and positivity.  The one-loop determinant, including these states, can be evaluated using heat kernel techniques \cite{Gaberdiel:2010xv}.  We note that the standard heat kernel technique, as used in \cite{Gaberdiel:2010xv}, implicitly contains within it a choice of boundary conditions.  By construction the  heat kernel computes a functional determinant on the space of linearized fields  obeying Dirichlet boundary conditions.  This  boundary condition allows for logarithmic modes.  So it is {not} the case that the computation of \cite{Gaberdiel:2010xv} implies that log gravity is correct and chiral gravity is incorrect.  It is just that the heat kernel of \cite{Gaberdiel:2010xv}, by construction, computes the one-loop determinant of log gravity rather than chiral gravity.  It would be interesting to construct an alternate version of the heat kernel which is appropriate for strict Brown-Henneaux boundary conditions; the resulting one-loop determinant would  agree with the answer obtained in  \cite{Maloney:2009ck} using algebraic methods.  This would presumably require the introduction of additional ghosts to  gauge  left-moving diffeomorphisms.

\section*{Acknowledgements}

We are grateful to M. Gaberdiel, A. Lepage-Jutier, J. Maldacena and R. Volpato.  This work is supported by the National Science and Engineering Research Council of Canada. TH is supported by U.S. Department of Energy grant DE-FG02-90ER40542.

\end{document}